\documentclass[aps,prl,superscriptaddress,reprint]{revtex4-1}
\usepackage{amsmath}
\usepackage{amssymb}
\usepackage{graphicx}
\usepackage{url}
\usepackage{hyperref}
\usepackage{color,xcolor}
\usepackage{epstopdf}
\usepackage{float}
\usepackage{ulem}

\begin{document}

\title{Quasi-one-dimensional ferroelectricity and piezoelectricity in WO$X_4$ halogens}
\author{Ling-Fang Lin}
\author{Yang Zhang}
\affiliation{Department of Physics and Astronomy, University of Tennessee, Knoxville, Tennessee 37996, USA}
\affiliation{School of Physics, Southeast University, Nanjing 211189, China}
\author{Adriana Moreo}
\author{Elbio Dagotto}
\affiliation{Department of Physics and Astronomy, University of Tennessee, Knoxville, Tennessee 37996, USA}
\affiliation{Materials Science and Technology Division, Oak Ridge National Laboratory, Oak Ridge, Tennessee 37831, USA}
\author{Shuai Dong}
\email{Corresponding author. Email: sdong@seu.edu.cn}
\affiliation{School of Physics, Southeast University, Nanjing 211189, China}
\date{\today}

\begin{abstract}
A series of oxytetrahalides WO$X_4$ ($X$: a halogen element) that form quasi-one-dimensional chains is investigated using first-principles calculations. The crystal structures, electronic structures, as well as ferroelectric and piezoelectric properties are discussed in detail. Group theory analysis shows that the ferroelectricity in this family originates from an unstable polar phonon mode $\Gamma_1^-$ induced by the W's $d^0$ orbital configuration. Their polarization magnitudes are found to be comparable to widely used ferroelectric perovskites. Because of its quasi-one-dimensional characteristics, the inter-chain domain wall energy density is low, leading to loosely-coupled ferroelectric chains. This is potentially beneficial for high density ferroelectric memories: we estimate that the upper-limit of memory density in these compounds could reach hundreds of terabytes per square inch.
\end{abstract}

\maketitle

\textit{Introduction.-}
Ferroelectrics with electric polarization ($P$) that can be switched in direction using weak external fields defines one of the most important branches of condensed matter physics and functional materials~\cite{Rabe:Bok,Scott:Jpcm,Scott:Science}. Their unique physical properties have already been widely applied in many commercial devices \cite{Scott:Science,Ahn:Science,scott2013ferroelectric}. For example, their ferroelectricity can be used to store information, as in non-volatile memories, while their piezoelectricity can transform forces/pressures into electric signals in sensors and micro-power sources \cite{Scott:Science,Ahn:Science,scott2013ferroelectric}. Moreover, their non-linear optical properties, pyroelectricity, as well as anomalous photocurrent effects are also of practical value~\cite{Auciello:pto}. Compared with conventional magnetic storages which need mechanically-suspended read/write heads, the pure electrical operations of ferroelectric (FE) memories are naturally preferred for better integration in devices. However, the typical currently available FE memory densities (typically 100 Mbits/inch$^2$) \cite{scott2013ferroelectric} are much lower than the magnetic ones (typically Tbits/inch$^2$) \cite{mallary2002one}. As a consequence, new materials with novel FE properties are much needed to fabricate high-density FE memories.

In recent years, two dimensional (2D) FE monolayers (or few layers) exfoliated from van der Waals (vdW) layered materials \cite{Wu:Wcms}, such as SnTe \cite{Chang:Science}, In$_2$Se$_3$ \cite{Ding:Nc} and CuInP$_2$S$_6$ \cite{Liu:Nc,You:Sa}, have established an emerging field due to its superiority at the nanoscale. For conventional FE perovskites with three dimensional (3D) pseudocubic structures, their ferroelectricity is often seriously suppressed in the very thin limit due to the depolarization field and broken bonds at surfaces, among other reasons \cite{Junquera:Nat,Batra:Prl,Zhong:Prl,Dawber:Rmp}. On the other hand, in the 2D FE materials mentioned above, the surfaces naturally form without dangling bonds, and thus a robust FE polarization can persist even in the very thin limit \cite{Chang:Science}.

As a consequence, it is natural to expect more exotic and better properties for applications in lower dimensional FE materials than in their 3D counterparts. In fact, the one-dimensional (1D) Poly (vinylidene fluoride-ran-trifluoroethylene) (PVDF-TrFE) FE nanowires can be easily fabricated, and they are predicted to have the potential for integration in devices with memory densities up to $33$ Gbits/inch$^2$ \cite{hu2009regular}. In addition, PVDF and its copolymers display exotic negative piezoelectricity \cite{Katsouras:Nm}. Recently, other 1D FE inorganic materials were also predicted to have interesting properties \cite{zjj:jacs2019}.

\begin{figure}
\centering
\includegraphics[width=0.48\textwidth]{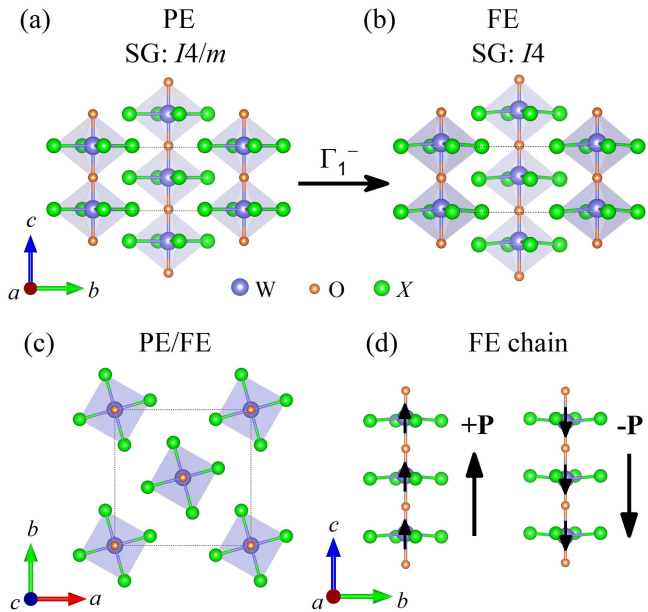}
\caption{Structures of WO$X_4$. Dash rectangles indicate the unit cells. (a-b) Side views of the PE and FE phases, respectively. Space groups (SG) are indicated.
The PE-FE phase transition is driven by the $\Gamma_1^-$ distortion,
which induces a net polarization $P$ along the $c$-axis. (c) Top view of the structure.
(d) Side view of the $P$ switch within a single chain.}
\label{structure}
\end{figure}

In this publication, a series of quasi-1D FE oxytetrahalides with the chemical formula WO$X_4$ ($X$ denoting a halide element: F, Cl, or Br) are theoretically investigated using Density Functional Theory (DFT) calculations. Among these materials, WOCl$_4$ and WOBr$_4$ have already been experimentally synthesized and their bulk crystal structures are shown in Fig.~\ref{structure} \cite{hess1966kristallstruktur,boorman1968mixed,muller1984wolframtetrabromidoxid,groh2013substitution}. Although there are no experimental reports for WOF$_4$ crystals yet (to our knowledge), preliminary DFT calculations are available \cite{osti_1295744}. As a summary of previous work, the members of the oxytetrahalides family were only briefly mentioned to be possible ferroelectrics, together with many other materials in the point group $4$  \cite{abrahams1999systematic}. However, no in-depth theoretical analysis has become available since those early studies. Our calculations fill this important gap and provide a systematic description of the FE properties of WO$X_4$,  as well as their piezoelectricity. Moreover, this work provides a natural extension of our recent effort addressing the 2D counterparts MO$_n$$X_{6-2n}$ with $n=2$ and M a group-VI transition metal~\cite{lin2019frustrated}.

\begin{table*}
\centering
\caption{Basic physical properties of the WO$X_4$ family from our DFT calculations (Cal.), compared with experimental results (Exp.) and with previous DFT studies. Data include: the lattice constants $a=b$ and $c$ (in units of \AA); volume $V$ (in units of \AA$^3$); characteristic bond length  W-O$1$ (shorter one), W-O$2$ (longer one), and W-$X$ (all in units of \AA); bond angles  $\angle$O-W-$X$ and $\angle$$X$-W-$X$ (in units of $^\circ$); band gap $E_{\rm g}$ (in units of eV); $P$ (in units of $\mu$C/cm$^2$); and energy barriers  $E_{\rm b1}$, $E_{\rm b2}$, and $E_{\rm bias}$ (in units of meV). SG:87 stands for Space Group 87. }
\begin{tabular*}{0.98\textwidth}{@{\extracolsep{\fill}}lcccccccccccccc}
\hline
\hline
Material                  &       & $a=b$ & $c$ & $V$   & W-O$1$& W-O$2$& W-$X$&$\angle$ O-W-$X$& $\angle$ $X$-W-$X$ & $E_{\rm g}$   & $P$  & $E_{\rm b1}$  & $E_{\rm b2}$& $E_{\rm bias}$\\
\hline
$X$ = F                  & Cal.  &6.55   &4.00 &171.57 & 1.74& 2.27&1.87  & 99.77        & 88.35 & 4.34&50.86&523&530&25\\
                         & Cal. \cite{osti_1295744}  &7.17   &3.93 &202.24 & 1.79& 2.15& 1.89 &  97.35       & 89.06 & 4.409&--&--&--&--\\
\hline
$X$ = Cl                 & Cal.  &8.57	 &3.93 &289.03 & 1.76&2.18 &2.30   &97.10          & 89.12 & 1.86&25.87&241&243&7\\
                         & Cal. \cite{osti_1206300}  &9.11   &3.91 &324.17 & 1.81& 2.10& 2.33 &  95.37       & 89.50 & 2.326&--&--&--&--\\
                         & Exp. \cite{hess1966kristallstruktur}  &8.48   &3.97 &285.35 & 1.74& 2.23& 2.28 &  98.49       & 88.75 & --&--&--&--&--\\
                         & Exp. \cite{groh2013substitution} &8.48   &4.00 &287.28 & 1.82& 2.18& 2.29 & 97.93        & 88.91 & --&--&--&--&--\\
\hline
$X$ = Br                 & Cal.  &9.17	 &3.86 &324.66 & 1.79&2.07 &2.47   &94.88          & 89.59 & 1.06&17.47&56&58&9\\
                         & Cal. \cite{osti_1206295} (SG:87)  &10.05   &3.84 &388.32 & 1.92& 1.92& 2.50 &  90.00       & 90.00 & 1.478&--&--&--&--\\
                         & Exp. \cite{muller1984wolframtetrabromidoxid} &9.00   &3.94 &318.88 & 1.78& 2.16&2.44  & 97.20        & 89.10 & --&--&--&--&--\\
\hline
\hline
\end{tabular*}
\label{Table1}
\end{table*}

\textit{DFT Methods.-}
In this work, first-principles DFT calculations were performed using the projector augmented wave (PAW) pseudopotentials with the Perdew-Burke-Ernzerhof (PBE) functional method as implemented in the Vienna {\it ab initio} Simulation Package (VASP) code \cite{Kresse:Prb99,Blochl:Prb2,Perdew:Prl08}. Because the $d^0$ configuration of W$^{6+}$ is not directly affected by the Hubbard $U$ repulsion, our results based on pure GGA, instead of GGA+$U$, should be sufficient. The lattice constants and atomic coordinates are fully optimized until the residual Hellman-Feynman forces become smaller than $0.01$ eV/{\AA}.

To better describe the inter-chain interaction, the vdW correction for potential energy was implemented by using the zero damping DFT-D3 method of Grimme~\cite{Grimme:Jcp}. The dynamical correlations between fluctuating charge distributions are incorporated in this method.

The Berry phase method was adopted to calculate $P$ \cite{King-Smith:Prb,Resta:Rmp}. The most likely ferroelectric switching pathways between different structures have been evaluated by using the nudged elastic band (NEB) method \cite{Henkelman:Jcp}.

In addition, the density-functional perturbation theory (DFPT) ~\cite{Gonze:prb} was employed to calculate the piezoelectric coefficients~\cite{Wu:PRB05}. The elastic matrix of WO$X_4$ has seven non-zero independent matrix elements ($C_{\rm 11}$, $C_{\rm 12}$, $C_{\rm 13}$, $C_{\rm 16}$, $C_{\rm 33}$, $C_{\rm 44}$, $C_{\rm 66}$) due to the $4/m$ class features of the $I4$ space group (No. 79)~\cite{Mouhat:PRB}. The piezoelectric tensor matrix has four independent matrix elements  ($e_{\rm 14}$, $e_{\rm 15}$, $e_{\rm 31}$, $e_{\rm 33}$) when the $4$ point group is considered~\cite{Jong:SD}.

\textit{Lattice properties.-}
As shown in Fig.~\ref{structure}, the structure of WO$X_4$ is formed by quasi-1D chains. Within these chains, each W ion is caged by an octahedron of four planar $X$'s and two apical O's, while the octahedra are corner-sharing via apical O's.

For the paraelectric (PE) structure [Fig.~\ref{structure}(a)], all W ions are at the center of the associated O$_2X_4$ octahedra, corresponding to the space groups $I4/m$ (No. $87$).

The $d^0$ configuration of W$^{6+}$ is FE active, as revealed in our recent studies of the 2D WO$_2X_2$ family of halogens \cite{lin2019frustrated}. The FE structure is shown in Fig.~\ref{structure}(b), corresponding to the space groups $I4$ (No. $79$). The positions of all W$^{6+}$ ions shift along the chain direction, leading to a net $P$. Here the AMPLIMODES software was employed to perform the group theoretical analysis, indicating that such a spontaneous distortion mode is the $\Gamma_1^-$ mode \cite{Orobengoa:Jac,Perez:Acsa}.

As shown in Fig.~\ref{structure}(c), the top view of the PE and FE phases are almost identical, implying that the FE chains are nearly isolated and only coupled via weak vdW interactions. Within each chain, the local distortion can be reversed directly, as shown in Fig.~\ref{structure}(d).

The optimized lattice structures of WO$X_4$ are summarized in Table~\ref{Table1}, and compared with experimental values and previous DFT results. Our calculated lattice constants are in excellent agreement with the experimental results for WOCl$_4$ and WOBr$_4$, while previous DFT results had large deviations \cite{osti_1295744,osti_1206300,osti_1206295}. There are several possible reasons for such discrepancy. First, a large value of $U$ ($=6.2$ eV) was applied on W in Refs.~\cite{osti_1295744,osti_1206300,osti_1206295}, while in our case $U$ is not included as explained before, because it is not necessary. Second, the vdW correction was not included in Refs.~\cite{osti_1295744,osti_1206300,osti_1206295}. This missing vdW interactions between chains cause the in-plane lattice constant to become seriously overestimated, namely more loose ($7-12\%$ larger). It is clear that the vdW interaction, as included in our calculation, is crucial to obtain the proper distances between chains.

Increasing the ionic size of $X^-$, the in-plane lattice constant ($a$=$b$), volume $V$, and bond length W-$X$ increase drastically. More specifically, the in-plane lattice constants increase by $31\%$ and $40\%$, and the volumes increase by $68\%$ and $89\%$ for $X$=Cl and Br, respectively, when compared with $X$=F. This huge in-plane expansion also reflects the softness induced by the vdW interactions. Note, for example that the lattice constants increase by $22\%$ and $29\%$ from NaF to NaCl and NaBr.

By contrast, the lengths of the W-O bonds and the lattice constant along the $c$-axis are nearly unchanged, or even slightly shorten, by increasing the ionic size of $X^-$. Moreover, the octahedra become increasingly flat, which represents a disadvantage for the FE mode (pseudo Jahn-Teller distortion). As a net result, the FE $P$ steadily decreases when the ionic size of $X^-$ increases, as discussed in the next subsection.

\begin{figure}
\centering
\includegraphics[width=0.48\textwidth]{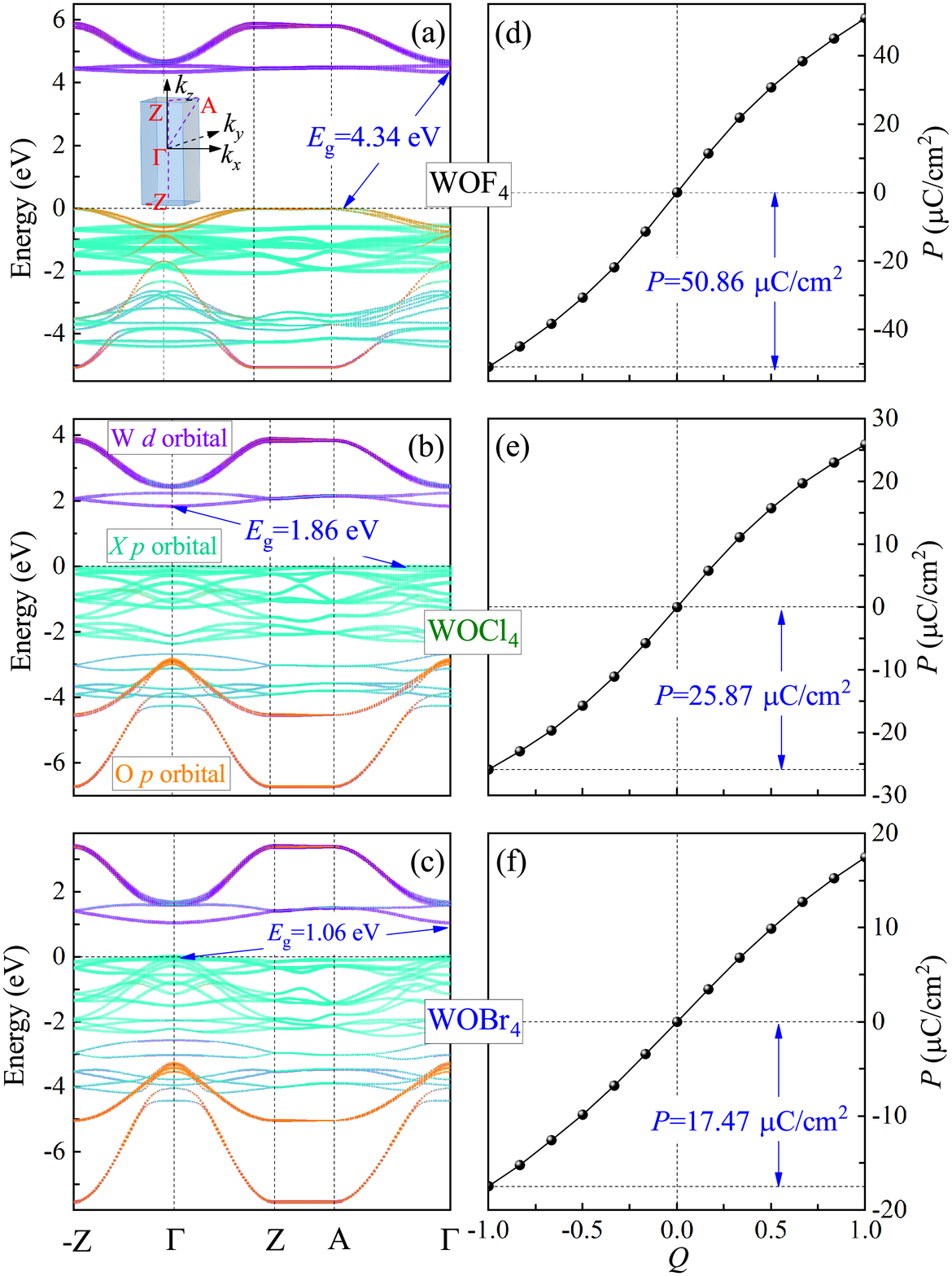}
\caption{Electronic structures of WO$X_4$. (a-c) Band structures. {\it Inset in (a)}: sketch of Brillouin Zones (BZ). (d-f) Estimation of the FE $P$ by changing the distortion amplitude ($Q$) of the $\Gamma_1^-$ mode continuously.}
\label{band}
\end{figure}

\begin{table*}
\centering
\caption{The calculated piezoelectric properties of the WO$X_4$ family, including elastic constants ($C_{ij}$'s, in units of GPa), piezoelectric stress coefficients ($e_{ij}$'s, in units of C/m$^2$), and piezoelectric strain coefficients ($d_{ij}$'s, in units of pC/N).}
\begin{tabular*}{0.98\textwidth}{@{\extracolsep{\fill}}lccccccccccccccc}
\hline
\hline
& $C_{\rm 11}$ & $C_{\rm 12}$ &$C_{\rm 13}$ & $C_{\rm 16}$ &$ C_{\rm 33}$ & $ C_{\rm 44}$ &$ C_{\rm 66}$ & $ e_{\rm 14}$ & $e_{\rm 15}$ & $e_{\rm 31}$ & $e_{\rm 33}$ & $d_{\rm 14}$ & $d_{\rm 15}$ & $d_{\rm 31}$ & $d_{\rm 33}$\\
\hline
WOF$_4$   & 14.6 & 10.3 & 8.4 & 2.4 & 88.3 & 13.8 & 12.5 & 0.010 & -0.045 & -0.062 &2.214 & 0.1 & -3.3 & -11.6 & 27.3\\
WOCl$_4$  & 15.4 & 10.6 & 7.1 & 0.3 & 61.9 & 7.0  & 10.7 & 0.004 & -0.029 & -0.029 &1.588 & 0.6 & -4.2 & -8.7  & 27.7\\
WOBr$_4$  & 15.3 & 8.7  & 6.8 & 0.5 & 69.6 & 5.5  & 11.6 & 0.002 & -0.034 & -0.042 &2.595 & 0.4 & -6.2 & -13.0 & 39.8\\
\hline
\end{tabular*}
\label{Table3}
\end{table*}

\textit{Electronic properties \& ferroelectricity.-}
The electronic structures of WO$X_4$ are shown in Figs.~\ref{band} (a-c). All three materials are indirect band gap insulators with moderate gap values ($E_{\rm g}$'s), which are beneficial for the measurements of ferroelectricity, although DFT calculations usually underestimate band gaps. The value of $E_{\rm g}$ decreases when the size of $X^-$ increases. The lowest conduction bands are all contributed by the W's $5d$ $d_{xy}$ orbital which is quite narrow, while the W's $d_{xz}$/$d_{yz}$ orbitals are slightly higher in energy and much broader. The topmost valence bands change from O's $2p$ dominance for $X$=F, to $2p$ orbital character for the cases of $X$=Cl and Br. This is reasonable since the electronegativity of the F element is the strongest, while the electronegativity of the O element is stronger than that of Cl or Br.

These electronic structures may be advantageous for optoelectronic applications due to the following characteristics. First, the valence bands originating in the $X$'s $2p$ orbitals are quite flat due to their isolated edge positions, which may be beneficial for optical absorption despite their indirect band gaps. Second, although the W's $d_{xy}$ band is narrow, their $d_{xz}$/$d_{yz}$ bands are much broader and they are only slightly higher in energy. For WOCl$_4$ and WOBr$_4$, photons above $1.8$ and $1.0$~eV can excite electrons to W's $d_{xz}$/$d_{yz}$ bands which have high mobilities along the $c$-axis, while the residual hole in the edge $X$'s $2p$ orbitals are much more localized. This aspect will be of interest for further investigations of optoelectronic currents using the materials proposed here.

Due to the drawbacks of DFT calculations when treating band gaps~\cite{zhang2000perspective}, these band gaps have been further checked using the hybrid functional calculation based on the HSE06 exchange (HSE)~\cite{Heyd:Jcp,Heyd:Jcp04,Heyd:Jcp06} and the Strongly Constrained Appropriately Normed (SCAN) meta-GGA semilocal exchange-correlation functional method~\cite{sun2015strongly}. As shown in Table S1 of the Supplementary Material (SM)~\cite{Supp}, the new calculated band gaps are only slightly larger than the values quoted in Table~I, showing that the main conclusions of the present work are not affected by using other functionals.

The values of $P$ are estimated using the Berry phase method \cite{King-Smith:Prb,Resta:Rmp}. Due to the Born-von Karman periodic boundary conditions, the possible values of $P$ for a fixed distortion can differ by multiples of the quantum $Q_{\rm p}$=$ec/V$, where $e$ is the charge of the electron, $c$ is the lattice constant in the direction of $P$, and $V$ is the volume of the unit cell. To avoid the ambiguity in the choice of the integer number for $Q_{\rm p}$, a continuous switching path of $P$ (from $+P$ to $-P$) is constructed by tuning the $\Gamma_1^-$ mode \cite{Neaton:Prb}, as displayed in Figs.~\ref{band}(d,e,f). Then, by this procedure the values of $P$ can be determined and they are $50.86$/$25.87$/$17.47$ $\mu$C/cm$^2$ for $X$=F/Cl/Br, respectively, which are comparable to traditional FE perovskites, such as BaTiO$_3$ ($\sim25$ $\mu$C/cm$^2$) \cite{Choi:Science}. Such a decreasing trend of the polarization $P$ agrees well with the aforementioned expectation arising from the structural flattening of the octahedra.

\begin{figure}
\centering
\includegraphics[width=0.48\textwidth]{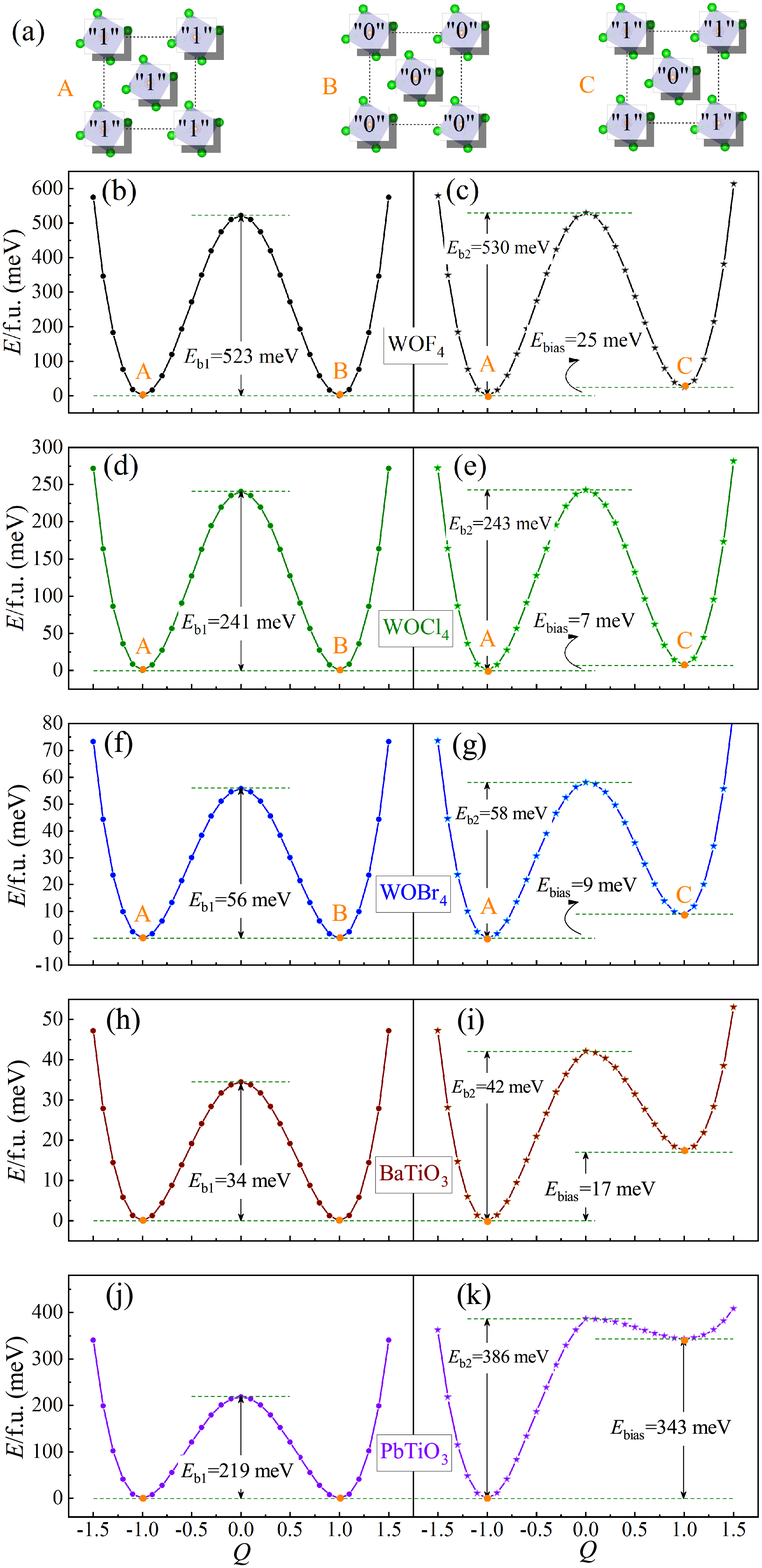}
\caption{(a) Top view of WO$X_4$ with different configurations. Logical states ``0" and ``1"
are used to indicate the different states ``+P" and ``-P" of each chain. (b-g) Energy barriers
for the WO$X_4$ family: (b,d,f) are when switching all chains together while (c,e,g) when switching half the chains. (h-k) Results for two classical FE materials, BaTiO$_3$ and PbTiO$_3$, were also computed for comparison, with a similar convention.}
\label{flip}
\end{figure}

The heights of the switching energy barriers are also important for potential applications of FE materials. Considering the weak vdW interaction between FE chains, here two simple and ideal paths are computed:
(a) switching all chains together and (b) switching only half the chains, indicated as A$\rightarrow$B and A$\rightarrow$C in Fig.~\ref{flip}(a), respectively. We found that reversing all chains at one time costs $523/241/56$ meV/f.u. for $X$=F/Cl/Br, as shown in Figs.~\ref{flip}(b,d,f), which is comparable to those of the two classical FE perovskites BaTiO$_3$ [$\sim34$ meV/f.u. in Fig.~\ref{flip}(h)] and PbTiO$_3$ [$\sim219$ meV/f.u. in Fig.~\ref{flip}(j)]~\cite{Cohen:Nat}. Then, we conclude that the binary coding in WO$X_4$ chains should be quite robust.

Comparing to the path A$\rightarrow$B, the energy barriers for path A$\rightarrow$C do not change much (they are only slightly higher), as indicated in Figs.~\ref{flip}(c,e,g). There is only a small energy bias $25/7/9$ meV/f.u. for $X$=F/Cl/Br between the structures A and C, implying that the vdW interactions between chains are weak and operations involving a single chain are feasible. For comparison, the flipping process of conventional FE perovskites BaTiO$_3$ and PbTiO$_3$ in a $\sqrt2\times\sqrt2\times1$ unit cell [corresponding to Fig.~\ref{flip}(a)] are also presented, as shown in Figs.~\ref{flip}(h-k). Then, the column by column flipping will significantly enhance the switching energy barriers. The most serious problem is that the state C is much higher in energy and as a consequence not robust, as presented in Figs.~\ref{flip}(i) and (k). Thus, the memory density based on conventional FE perovskites can not reach their atomic limit, while in principle our quasi-1D FE chains could. In other words, the minimal FE domains in 3D FE perovskites are large due to the high domain wall energies, while the minimal FE domains can be reduced to just one chain in the FE vdW materials described here due to the low domain wall energies. Thus, by considering each chain to be a minimal memory unit for one bit, the theoretical density upper limit can be easily estimated as $342$/$200$/$174$ Tb/inch$^2$ for $X$=F/Cl/Br, respectively.

In real situations, the switching process through domain wall motion can generally reduce the barrier. In fact, the A-C switching process can be thought as a domain-related one, because it involves the domain structure between chains and indeed changes the barriers. Due to the limitations of DFT, here only small size cells, already consisting of dozens of atoms, can be handled. This is not sufficient to include a real domain wall within chains, especially considering the fact that the inner-chain domain walls are charged (head-to-head or tail-to-tail) and thus quite high in energy. So the aforementioned energy barriers, calculated without the consideration of inner-chain domain walls, should be considered as upper limit values. The switching process including inner-chain domain wall deserves further computational (e.g. using the phase field model) and experimental studies.

Different from the simple A-B switching processes (reversing polarization of all chains simultaneously) which can be realized by applying a uniform electric field, it is a challenge to realize the A-C switching process in experiments due to the spatial resolution of those electric fields. Maybe it is possible to manipulate a single chain by using the tip of a scanning tunneling microscope. Here the A-C process is proposed as the minimum limit of ferroelectric switching. In real experiments, FE switching could be realized between the A-B and A-C limits, and gradually approach the A-C limit with advances in experimental techniques.

The double-well potential curves shown in Figs.~\ref{flip}(b), (d), and (f) can be fitted using the Landau-Ginzburg-Devonshire formula:
\begin{equation}
E=\frac{\alpha}{2}P^2+\frac{\beta}{4}P^4+\frac{\gamma}{6}P^6.
\label{eq1}
\end{equation}
Here, energy contributions up to the sixth order are necessary to describe the anharmonic double-well potential. The values of the energy coefficients in Eq.~\ref{eq1} are extracted from the DFT calculations, as listed in Table~\ref{Table2}.

\textit{Piezoelectricity.-}
In addition to the switchable polarization, there are important functional applications related to the piezoelectricity properties~\cite{Bellaiche:PRL00,bellaiche2002piezoelectricity,bellaiche1999intrinsic,Scott:Science,Fu:Nat00,Dawber:Rmp,Gou:AMI}. Recently, one-dimensional piezoelectric nanogenerators and other piezoelectric devices were widely developed~\cite{Qin:Nat08,Sun:NRM}. Here, the piezoelectric properties of WO$X_4$ can be evaluated by calculating the longitudinal piezoelectric coefficients.

As stated before, WO$X_4$ belongs to the non-centrosymmetric $I4$ space group where the elastic matrix has seven independent constants $C_{ij}$~\cite{Mouhat:PRB} and the piezoelectric stress tensor matrix has four independent coefficients $e_{ij}$ ~\cite{Jong:SD}. Then, the piezoelectric strain coefficient $d_{ij}$ can be calculated as:
\begin{equation}
d_{ij}=\sum_{k=1}^{6}e_{ik}S_{kj},
\end{equation}
where $S_{kj}$ is the elastic compliance coefficients ($S_{kj}$=$C_{kj}^{-1}$). Here, all piezoelectric coefficients are summarized in Table~\ref{Table3}.

As expected, the WO$X_4$ family has the largest elastic coefficients along the $c$-axis (i.e. $C_{\rm 33}$'s) due to the stacking of ionic bonds, while the other elements of the elastic matrix are much smaller than $C_{33}$ due to the vdW interactions. These values of the elastic matrix constants $C_{ij}$'s satisfy the Born stability criteria for the tetragonal $4/m$ class ~\cite{Mouhat:PRB}. As a consequence, WO$X_4$ ($X$=F, Cl and Br) should all be elastically stable even for the $X$=F case which is not experimentally reported to our knowledge. The key coefficients for piezoelectric performance $d_{\rm 33}$ are  $27.3$/$27.7$/$39.8$ pC/N for WO$X_4$ ($X$=F/Cl/Br), respectively. The diagonal elements of the piezoelectric stress tensor of the WO$X_4$ family are $2.214$/$1.588$/$2.595$ C/m$^2$, indicating that $P$ will be quite sensitive to stress. Both indicators $d_{33}$ and $e_{33}$ are larger than for the classic piezoelectric ZnO (with values $e_{33}$=$0.96$ C/m$^2$, $d_{33}$=$12.3$ pC/N)~\cite{Wu:PRB05} which is currently the most studied piezoelectric nanogenerator for low frequency applications \cite{wang2006piezoelectric,wang2004zinc}. In this framework, nanowires made of the WO$X_4$ family may be potentially useful as piezoelectric nanogenerators.

\begin{table}
\centering
\caption{The fitted Landau-Ginzburg-Devonshire parameters of Eq.~\ref{eq1} for WO$X_4$. Here, energies are in units of meV, polarizations $P$ in units of $\mu$C/cm$^2$, $\alpha$ in units of $10^{-2}$ meV/($\mu$C/cm$^2$)$^2$, $\beta$ in units of $10^{-4}$ meV/($\mu$C/cm$^2$)$^4$, and $\gamma$ in units of $10^{-6}$ meV/($\mu$C/cm$^2$)$^6$.}
\begin{tabular*}{0.48\textwidth}{@{\extracolsep{\fill}}lccc}
\hline
\hline
Material & $\alpha$ & $\beta$ & $\gamma$\\
\hline
$X$ = F & -48.82 &-1.81 & 0.14\\
\hline
$X$ = Cl &  -67.42 &-27.12 & 5.73\\
\hline
$X$ = Br &  -47.37 &-13.16 & 9.79\\
\hline
\hline
\end{tabular*}
\label{Table2}
\end{table}

\textit{Spin-orbit coupling \& temperature effects.-}
Note that all results above were obtained without the consideration of spin-orbit coupling (SOC). For the heavy element W, the SOC coefficient for its $5d$ orbitals can be considerable, which can cause a delicate Rashba splitting of conducting bands around the $\Gamma$ point in this polar system, as shown in Fig.~S1
in the SM~\cite{Supp}. However, the SOC effect to the structure-related ferroelectricity was found to be negligible since the $5d$ orbitals of W$^{6+}$ are nominally empty (see Table S2 in SM~\cite{Supp}).

In addition, {\it ab initio} molecular dynamics (MD) simulations have been employed to confirm if the ferroelectricity of WO$X_4$ persists to room temperature. Our MD results indicate that the ferroelectric order of WO$X_4$ chains indeed remains quite stable at $300$~K and even $400$~K (Figs. S2 and S3 in SM~\cite{Supp}), despite the existence of thermal fluctuations.

\textit{Conclusion.-}
In summary, using {\it ab initio} techniques in this work the WO$X_4$ family is predicted to be a novel series of ferroelectric materials displaying unusual quasi-one-dimensional characteristics. The robust ferroelectric distortion within each chain and weak vdW coupling between chains make them candidates for applications as high-density non-volatile memories at room temperature. Their electronic structures and piezoelectricity are also of potential value for optoelectronics and nanogenerators. Our predictions will  hopefully stimulate more theoretical and experimental works on transition metal oxyhalides, as well as other low-dimensional polar materials.

\acknowledgments{This work was supported by the National Natural Science Foundation of China (Grant Nos. 11834002 and 11674055). A.M. and E.D. were supported by the U.S. Department of Energy (DOE), Office of Science, Basic Energy Sciences (BES), Materials Science and Engineering Division. L.F.L. and Y.Z. were also supported by the China Scholarship Council.}

\bibliographystyle{apsrev4-1}
\bibliography{ref3}
\end{document}